\documentclass{jetpl}

\twocolumn
\usepackage{epstopdf}
\rus

\title{Chiral Optical Tamm States at the Interface between a Cholesteric and an All-Dielectric Polarization-Preserving Anisotropic Mirror}
\rtitle{Хиральное оптическое таммовское состояние \ldots}
\sodtitle{Хиральное оптическое таммовское состояние на границе холестерического жидкого кристалла и диэлектрического сохраняющего поляризацию анизотропного зеркала}

\author{Н. В. Рудакова$^{1}$\/\thanks{nrudakova@sfu-kras.ru},И.\,В.\,Тимофеев$^{1,2,3}$, 
Р.\,Г.\,Бикбаев$^{3,4}$, М.\,В.\,Пятнов$^{1,3}$, С.\,Я.\,Ветров$^{1,2}$, В.\,Ли$^5$\/\thanks{wlee@nctu.edu.tw}}
\rauthor{Н. В. Рудакова, И.\,В.\,Тимофеев, 
Р.\,Г.\,Бикбаев, М.\,В.\,Пятнов, С.\,Я.\,Ветров, В.\,Ли}
\sodauthor{Рудакова, Тимофеев, Бикбаев, Пятнов, Ветров, Ли}

\address{$^1$Институт инжинерной физики и радиоэлектроники, Сибирский Федеральный университет, Красноярск 660041, Россия\\
	$^2$Институт Физики им. Л.В. Киренского, Красноярский научный центр, СО РАН, Красноярск 660036, Россия\\
	$^3$Лаборатория нанотехнологий, спектроскопии и квантовой химии СФУ, 660041 Красноярск, Россия\\
	$^4$Политехнический институт, Сибирский Федеральный университет, Красноярск 660041, Россия\\
	$^5$Institute of Imaging and Biomedical Photonics, College of Photonics, National Chiao Tung University, Tainan 71150, Taiwan
}

\abstract{The chiral optical Tamm state is a new localized state of light at the interface between a polarization-preserving anisotropic mirror and an optically chiral medium such as a cholesteric liquid crystal. In this study the metal-free polarization-preserving mirror is used for efficient resonance control. We stress the advantage of the all-dielectric structure in obtaining high Q factor. The light is localized near the interface and the field decreases exponentially with the distance from the interface. The penetration of the field into the chiral medium is virtually blocked at wavelengths corresponding to the photonic band gap and close to the pitch of the helix. The polarization-preserving mirror has another photonic band gap as well. The energy transfer along the interface can be efficiently switched off by setting the tangential wave vector to zero. The spectral behavior of the chiral optical Tamm state is observed both as reflection and transmission resonance. This Fano resonance is due to circular polarizations interference and crossover, analogous to the Kopp–Genack effect. Our analytics agrees well with precise calculations, enabling intelligent design for laser and sensing applications.}


\begin{document}

\maketitle 

% \section*{Введение}
Электромагнитный аналог таммовского электронного состояния называется оптическим таммовским состоянием (ОТС), или, иначе, таммовским плазмон-поляритоном. В этом состоянии свет локализуется на общей границе двух сред, где происходит множественное отражение. ОТС нашли применение в различных оптических структурах, используемых для разработки новых способов и устройств управления светом [1, 2]. На границе холестерического жидкого кристалла (ХЖК) [3] и сохраняющего поляризацию анизотропного зеркала (СПАЗ) [4] возникает хиральное оптическое таммовское состояние (ХОТС) [5].

\begin{figure}[htbp]
\centerline{\includegraphics[width=8cm]{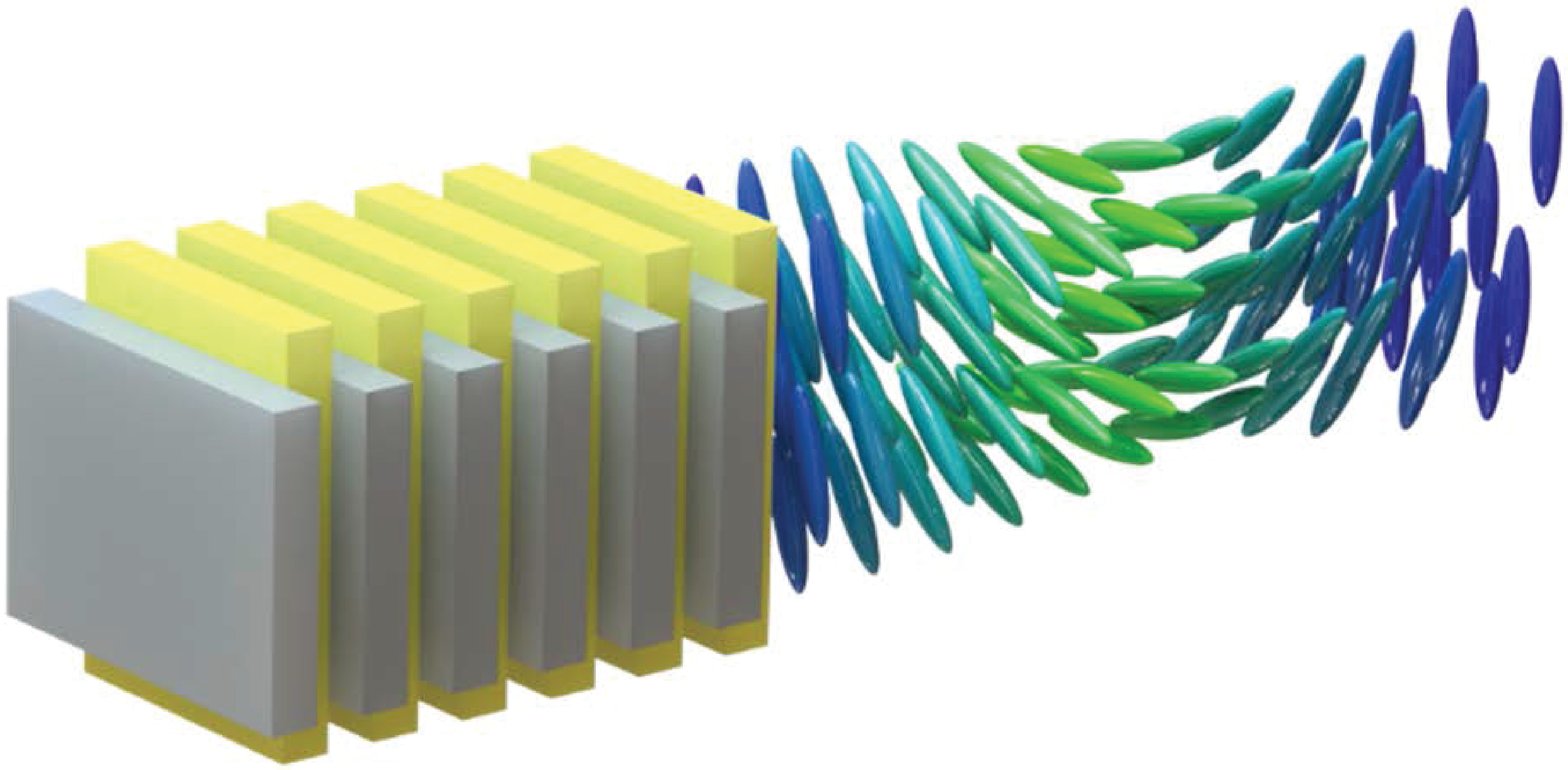}}
\caption{Рис.~1. Схематическое представление границы диэлектрического сохраняющего поляризацию анизотропного зеркала и холестерического жидкого кристалла}
\label{fig0}
\end{figure}

Многослойный СПАЗ представляет собой структуру, состоящую из чередующихся идентичных одноосные диэлектрических слоев с различными показателями преломления $n_e =\sqrt {\varepsilon _e } $ и $n_o =\sqrt {\varepsilon _o } $. Будем характеризовать его различными диэлектрическими тензорами двух соседних слоев $\hat {\varepsilon }_V $ и $\hat {\varepsilon }_H $. Число V-H пар (элементарных ячеек структуры) равно $N$, период структуры равен $\Lambda =2a$, где $a$ -- толщина одного слоя. Холестерический жидкий кристалл -- это оптическая хиральная среда, обладающая непрерывной винтовой симметрией тензора диэлектрической проницаемости. Для него вводятся следующие характеристики: величина шага спирали $p~=$~1~мкм, $L$ -- толщина слоя холестерика, показатели преломления обыкновенного и необыкновенного лучей $n_{\left\| \right.} =\sqrt {\varepsilon _{\left\| \right.} } $ и $n_\bot =\sqrt {\varepsilon _\bot } $ близки по значениям к $n_e $ и $n_o $ для СПАЗ.

\begin{figure}[htbp]
\centerline{\includegraphics[width=8cm]{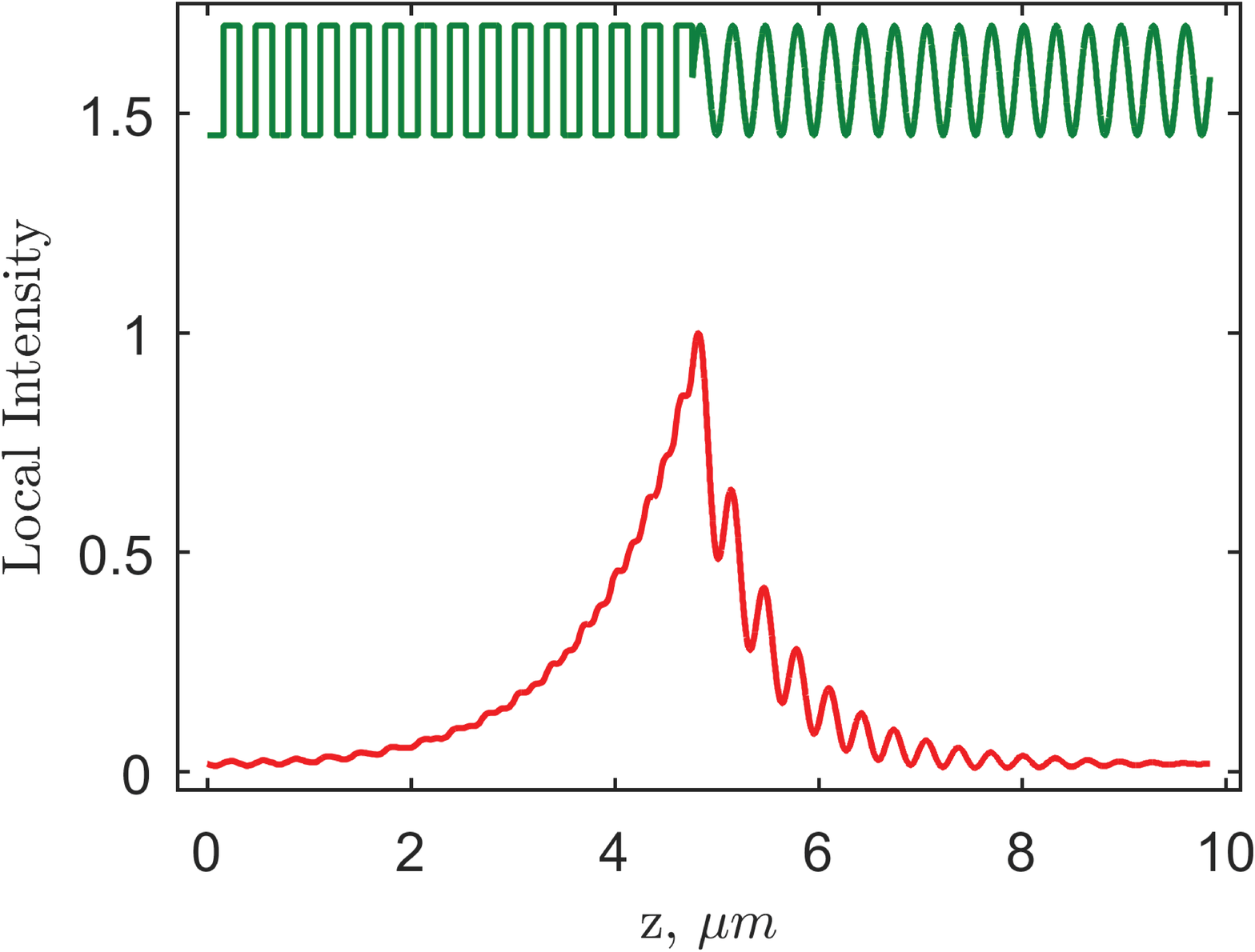}}
\centerline{\includegraphics[width=8cm]{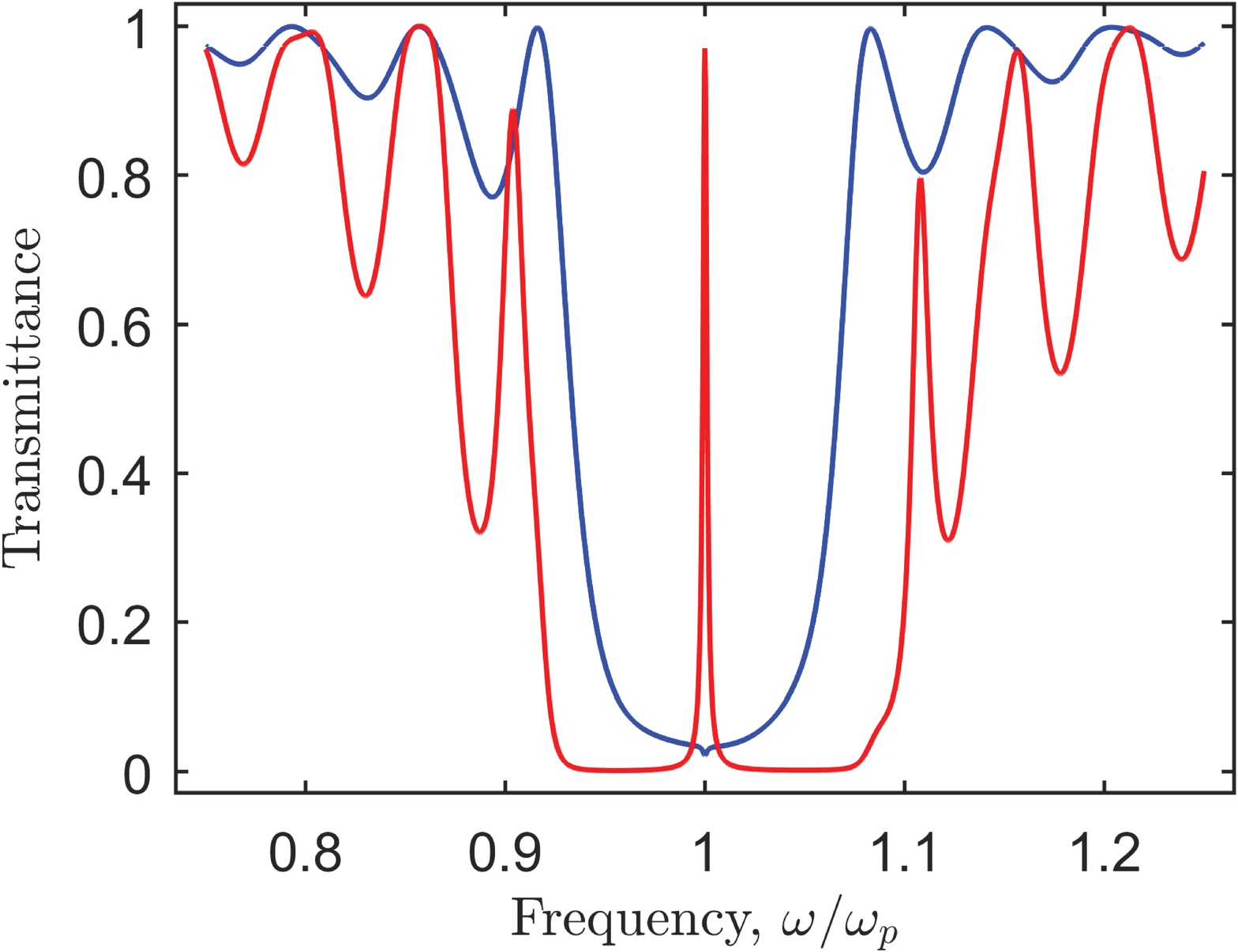}}
\caption{Рис.~2. а) Эффективный показатель преломления (зеленый) в зависимости от глубины $z$ слоистой структуры и нормированная локальная интенсивность $\vert $Е$\vert ^{2}$ хирального оптического таммовского состояния (красный). б) Спектры пропускания для круговых поляризаций, совпадающей (красный) и противоположной (синий) по отношению к холестерику. Показатели преломления для СПАЗ и ХЖК совпадают и соответствуют значениям для нематических жидких кристаллов семейства цианобифенилов: $n_{o}~=~$1.45, $n_{e}~=~$1.7. Длина волны в центре запрещенных зон обоих зеркал составляет $\lambda _{c}$~=~$\lambda _{p}~=$~1 мкм. СПАЗ содержит 30 слоев с общей толщиной 30/2($n_{o}~+~n_{e})~\approx ~$4.76 мкм. Слой ХЖК содержит 8 шагов спирали с общей толщиной 8$\cdot $2/($n_{o}~+~n_{e})~\approx ~$5.08 мкм.}
\label{fig1}
\end{figure}

СПАЗ является периодической слоистой средой, которую можно считать брегговским отражателем, тогда его рассмотрение можно проводить с помощью аналитических формул [6]. Выражение для коэффициента отражения:
\begin{equation}
\label{eq1}
r_N =\frac{CU_{N-1} }{AU_{N-1} -U_{N-2} }=\frac{C}{A-\frac{\sin \left( {N-1} \right)K\Lambda }{\sin NK\Lambda }},
\arg r_N =\theta ,
\end{equation}
где $U_N =\frac{\sin \left( {N-1} \right)K\Lambda }{\sin K\Lambda },$ а элементы матрицы преобразования для одной ячейки, связывающие амплитуды плоских волн в первом слое с аналогинными амплитудами 
%для первого слоя 
в соседней элементарной ячейке, выглядят следующим образом:
\begin{eqnarray}
\label{eq2}
A=e^{ik_{1z} a}\left[ {\cos k_{2z} a+\frac{1}{2}i\left( {q+\frac{1}{q}} \right)\sin k_{2z} a} \right]; \\
C=e^{ik_{1z} a}\left[ {-\frac{1}{2}i\left( {q-\frac{1}{q}} \right)\sin k_{2z} a} \right].
\end{eqnarray}
Здесь $k_{1z} =\left( {\omega /c} \right)n_e , \quad k_{2z} =\left( {\omega /c} \right)n_o $ -- волновые вектора, соответственно для первой и второй сред, ${k_{2z} } \mathord{\left/ {\vphantom {{k_{2z} } {k_{1z} =q}}} \right. \kern-\nulldelimiterspace} {k_{1z} =q}$, блоховское волновое число дается выражением $\cos K\Lambda =ReA.$

Коэффициент отражения для холестерика представим в следующем виде [3]:
\begin{eqnarray}
\label{eq3}
r_L =\frac{i\delta \sin \beta L}{\frac{\beta \tau }{\kappa ^2}\cos \beta L+i\left[ {\left( {\frac{\tau }{2\kappa }} \right)^2+\left( {\frac{\beta }{\kappa }} \right)^2-1} \right]\sin \beta L},
\\
\arg r_L =\chi ,
\end{eqnarray}
здесь $\beta =\kappa \left[ {1+\left( {\frac{\tau }{2\kappa }} \right)^2-\left[ {\left( {\frac{\tau }{\kappa }} \right)^2+\delta ^2} \right]^{1/2}} \right]^{1/2}$, $\bar {\varepsilon }=\frac{\varepsilon _\bot +\varepsilon _{\left\| \right.} }{2},\,\,\,\,\,\,\delta =\frac{\varepsilon _{\left\| \right.} -\varepsilon _\bot }{\varepsilon _\bot +\varepsilon _{\left\| \right.} },\,\,\,\,\kappa =\frac{\omega \bar {\varepsilon }}{c},\,\,\,\,\tau =\frac{4\pi }{p}$. Уравнение на фазу $\chi $ отраженной от ХЖК волны для полубесконечного ХЖК может быть записано в виде [5]:
\begin{equation}
\label{eq4}
\cos 2\chi =\frac{\lambda -\lambda _c }{\Delta \lambda },
\end{equation}
здесь $\lambda_c = 2\pi c / \omega$ -- длина волны в центре запрещенной зоны ХЖК, 
$\Delta \lambda =\lambda _c \delta /\bar {\varepsilon }$ -- ширина запрещенной зоны ХЖК. В этих же терминах для СПАЗ также можно приближенно записать уравнение на фазу $\theta $ отраженной от СПАЗ волны:
\begin{equation}
\label{eq5}
\sin 2\theta \approx -\frac{\omega -\omega _p }{\Delta \omega }={\left( {\frac{\omega \Lambda }{\pi c}-1} \right)} \mathord{\left/ {\vphantom {{\left( {\frac{\omega \Lambda }{\pi c}-1} \right)} {\frac{\pi }{2}\left( {\frac{1-q}{1+q}} \right)}}} \right. \kern-\nulldelimiterspace} {\frac{\pi }{2}\left( {\frac{1-q}{1+q}} \right)},
\end{equation}
здесь $\omega _p ={\pi c} \mathord{\left/ {\vphantom {{\pi c} \Lambda }} \right. \kern-\nulldelimiterspace} \Lambda $ -- частота в центре запрещенной зоны СПАЗ.

Запишем выражение для локализации поля на границе структуры СПАЗ-ХЖК, используя формулы (\ref{eq4}) и (\ref{eq5}), а также учитывая условия сшивки полей на границе раздела сред:
\begin{equation}
\label{eq6}
\pi N=\theta +\chi +\varphi .
\end{equation}
Здесь $\varphi $ -- угол между оптическими осями ХЖК и СПАЗ на границе раздела. За цикл из двух переотражений набегает  геометрическая составляющая фазы $2\varphi$.

На рисунке \ref{fig1}a с помощью эффективного показателя преломления показана структура, состоящая из холестерического жидкого кристалла и СПАЗ в зависимости от глубины $z$ слоистой структуры, а также нормированная локальная интенсивность $\vert $Е$\vert ^{2 }$для ХОТС. Из рисунка видно, что свет локализован около границы СПАЗ-ХЖК, а локальная интенсивность поля спадает экспоненциально с увеличением расстояния от этой границы. Рисунок 2б представляет спектры пропускания для двух противоположных круговых поляризаций. Для поляризации, знак которой совпадает со знаком закручивания спирали холестерика, пропускание имеет резонансный пик в центре запрещенной зоны, тогда как для поляризации противоположного знака пик отсутствует.

На рисунке 3 представлены спектры пропускания структуры, рассчитанные при помощи метода Берремана для различных положений запрещенных зон ХЖК и СПАЗ в нормированных частотах. При равенстве анизотропий запрещенная зона для ХЖК шире, чем для СПАЗ, это преимущество ХЖК-зеркала компенсируется тем, что оно отражает лишь одну круговую поляризацию. Видно, что в области частот, где запрещенные зоны ХЖК и СПАЗ полностью либо частично перекрываются, появляются резонансные пики пропускания. Их положение качественно согласуется с расчетом по уравнению (\ref{eq6}). Количественное отличие от точного расчета проявляется вблизи угла $\varphi =-\pi /4$, когда теория предсказывает переход ХОТС в краевую моду, выход резонансного пика на край запрещенной зоны и появление пика на противоположном краю запрещенной зоны. При фиксированном $\varphi =-\pi /4$ пик наблюдается на коротковолновом краю запрещенной зоны при $\omega _c /\omega _p >1$ и на длинноволновом краю запрещенной зоны при $\omega _c /\omega _p <1$. На нижнем графике рисунка 3 приведен расчет для случая, когда толщины обоих зеркал увеличены в 10 раз. Решения уравнений (1-3) для конечных сред, изображенные красным цветов, приближаются к решениям уравнений (4,5) для полубесконечных сред, изображенным синим цветом. Приближенное уравнение (\ref{eq5}) дает видимое различие решений. Это различие становится сколь угодно малым при стремлении нормированной анизотропии $\delta /\bar {\varepsilon }$ к нулю.

\begin{figure}[htbp]
\centerline{\includegraphics[width=8cm]{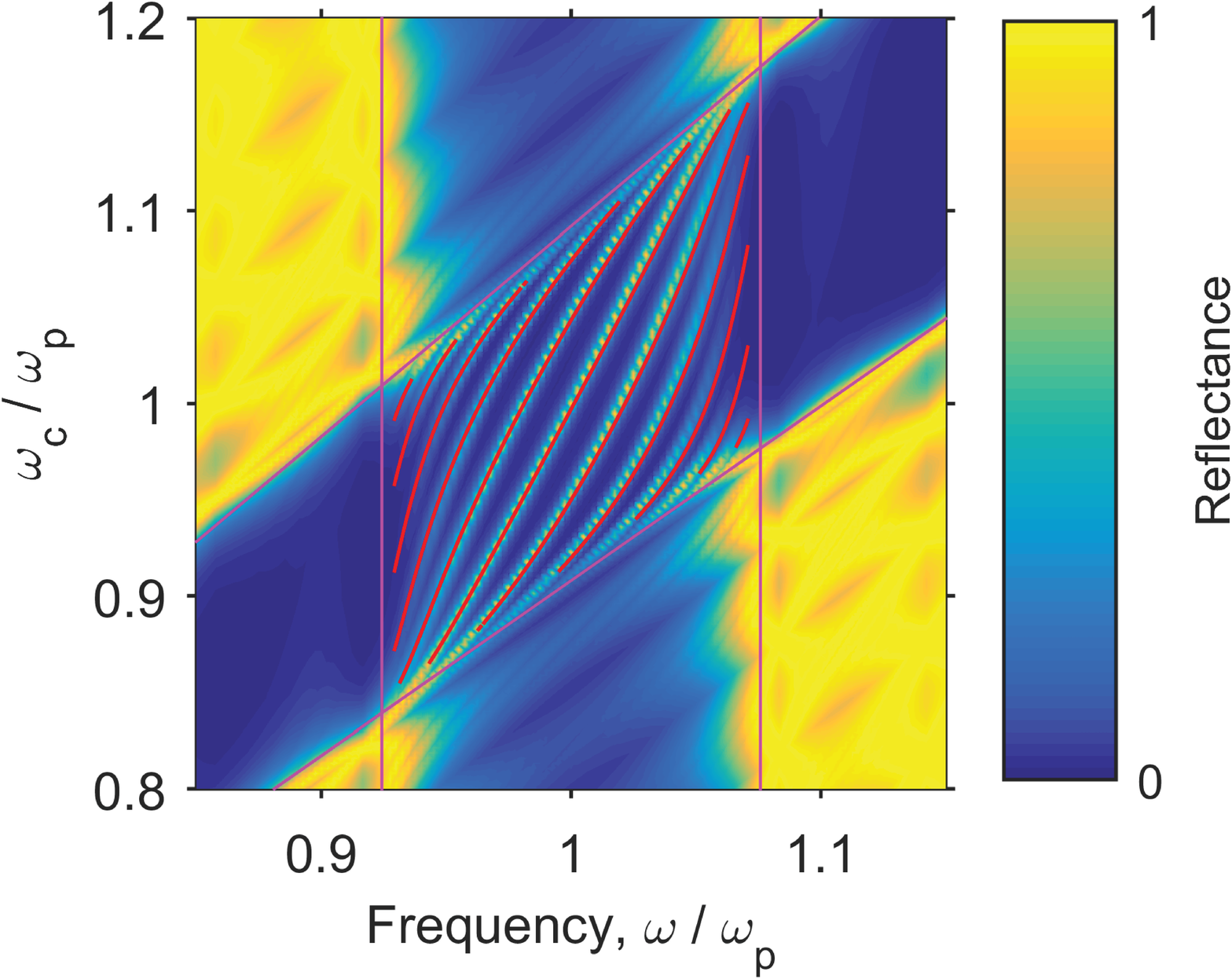}}
\centerline{\includegraphics[width=8cm]{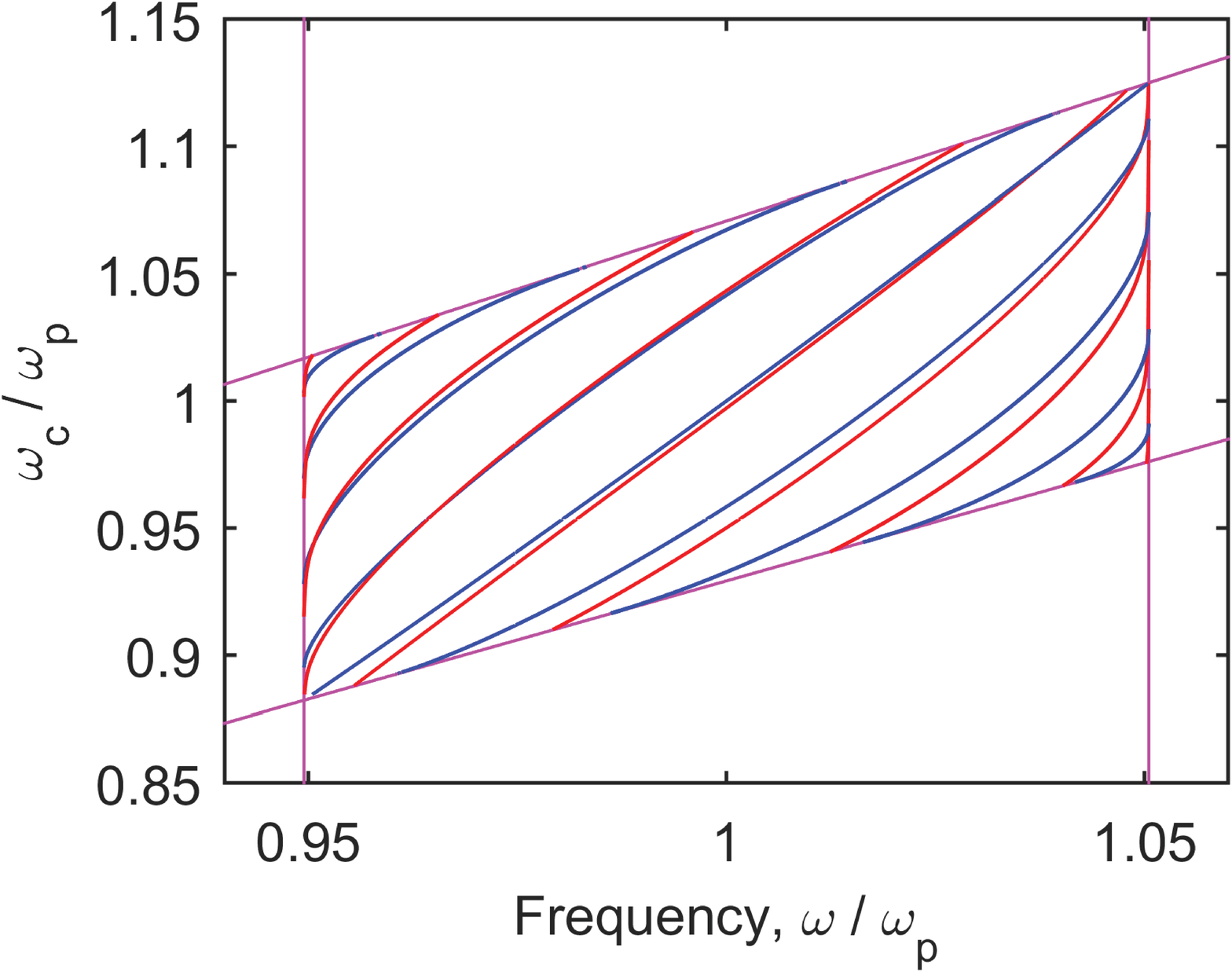}}
\caption{Рис.~3. Спектр пропускания при изменении шага спирали ХЖК и смещении частоты центра его запрещенной зоны $\omega _c $. Частота центра запрещенной зоны СПАЗ -- $\omega _p $. При $\omega _c /\omega _p =1$ параметры соответствуют рисунку 1. Угол между оптическими осями ХЖК и СПАЗ на границе раздела пробегает значения в интервале $\pi /4\leqslant \varphi \leqslant 5\pi /4$ с шагом $\pi /10$. Прямые линии пурпурного цвета указывают границы запрещенных зон. В области перекрытия запрещенных зон приведены решения уравнения (\ref{eq6}), с учетом уравнений (1-3) для конечных сред -- красным цветов, с учетом уравнений (4,5) для полубесконечных сред -- синим цветом.}
\label{fig4}
\end{figure}

Коэффициент пропускания в пиках не достигает единицы, как и на рисунке 2б. Это объясняется различием зеркал и отсутствием пространственной симметрии. На языке временной теории связанных мод [7] это называется нарушением условия критической связи. Точнее говоря, рассматриваются два основных канала релаксации ХОТС: через СПАЗ и через ХЖК. Толщина ХЖК выбирается достаточно большой, чтобы проходящее излучение имело круговую поляризацию, знак которой противоположен знаку поляризации ХОТС. Толщина СПАЗ подбирается так, чтобы выровнять скорости релаксации в каналах. Погрешность объясняется отличием поляризации от круговой и пропорциональна нормированной анизотропии $\delta /\bar {\varepsilon }$.

Исследовано хиральное оптическое таммовское состояние на границе холестерического жидкого кристалла и слоистой структуры, характеризующейся чередующимися одинаковыми одноосными диэлектрическими слоями с ортогональными направлениями оптических осей. Показано, что найденное состояние является высокодобротным, а также может быть эффективно перестроено по частоте.

\begin{center}
ЛИТЕРАТУРА
\end{center}

1.~A.V. Kavokin et al. // Phys. Rev. B. 2005. V. 72. P. 233102.

2.~M. Kaliteevski et al. // Phys. Rev. B. 2007. V. 76. P. 165415.

3.~В.А. Беляков, А.С. Сонин ``Оптика холестерических жидких кристаллов'' М: Наука, 1982.

4.~N.V Rudakova et al. // Bull. Russ. Acad. Sci. Phys. 2017. V. 81. P. 10.

5.~И.В. Тимофеев, С.Я. Ветров, // Письма в ЖЭТФ 2016 Т. 104, В. 6, С. 393--397.

6.~А. Ярив, П. Юх ``Оптические волны в кристаллах'' M.: Мир, 1987.

7.~J.D. Joannopoulos, S.G. Johnson, J.N. Winn, R.D. Meade ``Photonic Crystals Molding the Flow of Light'' 2nd edition. Princeton Univ. Press, 2008.

\end{document}